\documentclass[letterpaper, 10 pt, conference]{ieeeconf}  

\IEEEoverridecommandlockouts                              

\usepackage{graphicx}
\graphicspath{{Figures/}}

\usepackage{mathtools,amsmath,amssymb,amsfonts} 
\usepackage{wrapfig}
\usepackage{epsfig,epstopdf}
\usepackage{color,colortbl}
\usepackage{algorithm}
\usepackage[version=4]{mhchem}
\usepackage{siunitx}
\usepackage{array}
\usepackage{longtable}
\usepackage{cleveref}

\usepackage{empheq}

 \newcommand\umax{u_\text{M}}

\definecolor{dgreen}{rgb}{0.0, 0.42, 0.24}

\DeclareMathOperator{\sat}{sat}

\newtheorem{theorem}{Theorem}
\newtheorem{remark}{Remark}
\title{\LARGE \bf
A hybrid dynamical system approach to the impulsive control of spacecraft rendezvous}

\author{Alexandre Seuret $^{1}$, Rafael Vazquez $^{1}$ and Luca Zaccarian$^{2}$
\thanks{$^{1}$Alexandre Seuret and Rafael Vazquez are with the University of Sevilla, Sevilla Spain.
        {\tt\small aseuret,rvazquez1@us.es}}%
\thanks{$^{2}$Luca Zaccarian is with the LAAS-CNRS, Toulouse, France and with the University of Trento, Trento, Italy
        {\tt\small zaccarian@laas.fr}}%
}

\begin{document}

\maketitle

\begin{abstract}
This paper introduces a hybrid dynamical system methodology for managing impulsive control in spacecraft rendezvous and proximity operations under the Hill-Clohessy-Wiltshire model. We address the control design problem by isolating the out-of-plane from the in-plane dynamics and present a feedback control law for each of them. This law is based on a Lyapunov function tailored to each of the dynamics, capable of addressing thruster saturation and also a minimum impulse bit. These Lyapunov functions were found by reformulating the system's dynamics into coordinates that more intuitively represent their physical behavior. The effectiveness of our control laws is then shown through numerical simulation.
\end{abstract}

\section{Introduction}
From the increasing complexity of space missions, emerged the needs of servicing satellites effectively. Operations such as inspection, repair, refueling, and monitoring are essential, requiring a spacecraft, known as the chaser, to execute precise maneuvers near a target spacecraft. These maneuvers, known as rendezvous and proximity operations, are critical  in guiding a spacecraft to a pre-determined proximity to the target to perform mission-specific tasks. The demand for autonomous guidance and control in these tasks is more pressing than ever, motivated by key space activities like asteroid mining~\cite{Hein20}, collision avoidance~\cite{Lee18}, on-orbit assembly~\cite{Underwood15}, debris removal~\cite{Sasaki19}, and resupplying missions~\cite{Souza07}. Originating from the Apollo program's lunar orbital rendezvous concept~\cite{Neufeld08}, which was essential for reducing payload mass and the feasibility of the mission, the techniques have considerably developed. Proximity operations have since become commonplace, such as in the frequent rendezvous missions to the International Space Station, and continue to be integral in the advancement of low Earth orbit operations and beyond. The most basic rendezvous model is described by the Clohessy-Wiltshire (HCW) equations~\cite{CW}, which were in fact developed for the Apollo program and consider the target in a circular Keplerian orbit and the chaser in close proximity. 

In the past, many works have addressed the design of feedback laws for the rendezvous problem by using a Model Predictive Control (MPC) strategy, see e.g.~\cite{GAVILAN2012111,FlatnessBased,Vaz17,San20a}. MPC can contribute important properties such as  optimality and constraint handling, however at a considerable computational cost that may not be possible e.g. onboard low-cost satellites such as Cubesats; in addition it is not a trivial task to provide guarantees on stability and feasibility with MPC, or to handle minimum impulse bits. On the other hand, despite the potential of hybrid systems theory~\cite{goebel2012hybrid} for ensuring stability, simplifying calculations, and improving efficiency, it has rarely been applied to the rendezvous problem, with limited exceptions such as~\cite{brentari2018hybrid} (and references therein). Our approach diverges from these instances by utilizing a simpler simulation model but incorporating saturations into our analysis, and avoiding the need for an optimization process in the control law computation.

This paper considers the terminal rendezvous stage  with a focus on designing efficient impulsive maneuvers. Traditional planning for such missions simplifies the process by approximating actual maneuvers with instantaneous velocity changes, a widely accepted practice that justifies the impulsive approach we use in this paper. Within a hybrid systems framework~\cite{goebel2012hybrid}, we separate the out-of-plane  and in-plane dynamics and propose a feedback control law based on a specific Lyapunov functions for each of the dynamics that is able to both thruster saturation and minimum impulse bit. These Lyapunov functions are found by expressing the dynamics in more natural coordinates that capture their physical behaviour. Finally, our control law's performance is validated through simulations. 

The structure of the manuscript is as follows: Section~\ref{sec-rendezvous} introduces the HCW model used in our approach. Section~\ref{sec-outofplane} outlines the hybrid stabilization for the out-of-plane dynamics whereas Section~\ref{sec-inplane} similarly deals with the stabilization of the in-plane dynamics. 
The different stabilizers are combined in a unique feedback law whose properties are established in Section~\ref{sec:combined_feedback}.
Next, Section~\ref{sec-simulations} provides simulations results, and the paper is closed in Section~\ref{sec-concl} with some concluding remarks.

\section{Problem Statement}\label{sec-rendezvous}
There are numerous mathematical models for
spacecraft rendezvous.  if the target is orbiting in a \emph{circular}
Keplerian  orbit and the approaching vehicle (chaser) is close to the
target, then the linear Hill-Clohessy-Wiltshire (HCW) equations,
introduced in~\cite{Hill} and~\cite{CW}, describe with adequate
precision the relative position of the spacecraft.

\subsection{HCW model}
The HCW model assumes that the target vehicle is passive and
moving along a circular orbit. Using impulsive control, it describes the relative motion of a chaser vehicle close to the target and can be formulated as the following equation.
\begin{subequations}\label{HCW}
\begin{align}
    \Ddot{r}_x - 3n^2 r_x  - 2 n  \dot{r}_y & = 0, \label{eq:HCW-a} \\
     \Ddot{r}_y + 2n \dot{r}_x & = 0 ,\label{eq:HCW-b} \\
     \Ddot{r}_z + n^2  r_z & = 0 ,\label{eq:HCW-c}
\end{align}
\end{subequations}
where
\begin{itemize}
    \item $r=(r_x,r_y,r_z)\in\mathbb R^3$ stands for the relative position between the chaser and the target in the target reference frame.
    \item To avoid any confusion, the time derivative of $(r_x,r_y,r_z)$  will be denoted as $v=(v_x, v_y, v_z)$, which stands for the relative velocities between the chaser and the target in the target reference frame.  
    \item $n$ as defined before is the mean orbital angular speed of the target, which in this case (being the target's orbit circular) coincides with its instantaneous angular speed.  The angular speed of the target through its orbit is $n=\sqrt{\frac{\mu}{R^3}}$, where $\mu$ is the gravitation
parameter of the Earth, $\mu=398600.4~\mathrm{km^3/s^2}$ and $R$ is the radius. Thus for a typical orbit at, say, an altitude of 500 kilometers we would get $n=0.0011$ rad/s.
\end{itemize}

\begin{figure}
\begin{center}
\includegraphics[width=0.8\columnwidth]{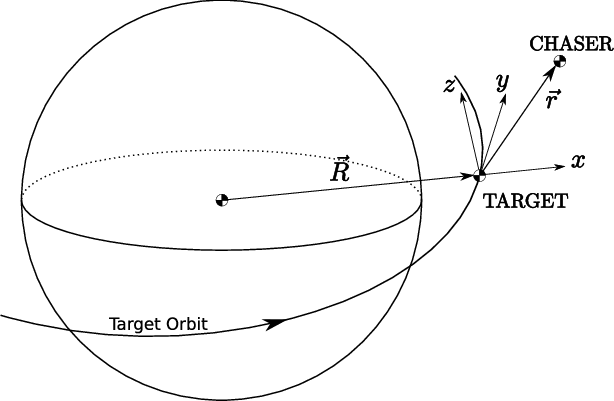}    
\caption{Local-Vertical, Local-Horizontal (LVLH) frame.}
\label{fig-lvlh frame}
\end{center}
\end{figure}

Thus the full state is characterized by $(r,v)\in\mathbb R^6$. Equations \eqref{HCW} are expressed in a local target reference frame (LVLH), which is a rotating frame centered at the target (see Figure~\ref{fig-lvlh frame}). The coordinates lend themselves to a physical explanation: $r_x$ is the radial distance (with a positive $r_x$, the chaser is at a higher altitude than the target, with a negative $r_x$ it is at a lower altitude); $r_y$ is the along-track distance (the phase with respect to the target's orbit). Thus, $r_x=r_z=0$ and $r_y\neq 0$ are equilibria, representing the chaser in the same orbit as the target but with some phase lag: delayed with respect to it or ahead with respect to it. The $r_x-r_y$ motion is coupled because, due to the laws of orbital mechanics, if the chaser is further away from Earth (higher values of $r_x$) it should move slower than the target and accumulate a delay in the $r_y$ direction (drifting back). If chaser is closer to the Earth (smaller values of $r_x$), then it moves faster than the target and it surpasses it, drifting ahead. These drifts are very slow, thus the dynamics are weakly unstable (as represented by a double eigenvalue at zero).

Finally, $r_z$ is the relative vertical distance to the target's orbit, describing how  the plane of the chaser's orbit is related to the orbital plane of the target. Physically, the $r_z$ motion is decoupled from the other ones because orbital mechanics dictates that orbital motions stay in a plane; under the HCW approximation, nothing that one can do in the $r_x-r_y$ plane can change this plane, thus $r_z$ remains decoupled, and vice-versa: changing the plane does not affect what happens in $r_x-r_y$. There is no drift in $r_z$ as seen in the equations, since the chaser's orbital plane  remains the same if one does not act on this direction.

\subsection{Propulsive model and constraints}

In this paper, we consider that the control action is performed as impulses acting only on the velocities, that is, at specific instants decided by the control law, the velocities experience the following discontinuous motion
\begin{subequations}\label{eq:ImpCont}
\begin{align}
v_x^+ = v_x +u_x,\label{eq:ImpContx}\\
    v_y^+ = v_y +u_y,\label{eq:ImpConty}\\
    v_z^+ = {v_z} +u_{z},\label{eq:ImpContz}
    \end{align}
\end{subequations}
where we defined the control input $u=(u_x, u_y, u_z)\in \mathbb R^3$.

Model \eqref{HCW},\eqref{eq:ImpCont} shows that the dynamics of a spacecraft rendezvous are governed by open-loop continuous-time dynamics. The only control action occurs at some time instants. It consists in an abrupt change of the velocity in all the directions of the target reference frame.
Regarding the possible values of thrust (namely of $u$), we assume that the control inputs are bounded in absolute value. This means that there exists $\umax>0$ such that
\begin{equation} \label{eqn-uSAT}
\begin{array}{lcl}
 |u_{x}| \leq \umax,\quad |u_{y}| \leq \umax,\quad |u_{z}| \leq \umax,
\end{array}
\end{equation}

We assume that $|u_{(\cdot)}|$ can take any value in the interval, i.e., it is assumed that thrusters valves can be opened partially to produce the exact amount of force commanded by the control law.  Thus, we define the symmetric saturation nonlinear map $\sat$, whose components are defined as follows
\begin{equation}
\label{eq:nonlinearity}
    \sat_i(u_i) = \begin{cases}
    \mathrm{sign}(u_i) \umax ,&  \mbox{ if } |u_i| > \umax, \\
    u_i,                  &  \mbox{ if } 0 \leq |u_i| \leq \umax. \\
    \end{cases}
\end{equation}
\begin{remark}
In practice, the thrusters may also be bounded from below, i.e.  control inputs that are too small cannot be produced by the thrusters. This issue will not be considered in this paper and is left as future work. 
\end{remark}


\subsection{Control objectives}\label{sec-obj}
\label{control_section}

In view of the hybrid nature of the spacecraft rendezvous with mixed continuous and discrete dynamics, we propose here to use the hybrid dynamical systems theory \cite{goebel2009hybrid,goebel2012hybrid}, to propose nonlinear hybrid control laws for system \eqref{HCW}-\eqref{eq:ImpCont} accounting for the input constraints \eqref{eq:nonlinearity}.


The idea to deploy the Hybrid Dynamical Systems framework on spacecraft rendezvous is not new since it has already been adopted in previous research \cite{brentari2018hybrid, brentari2019hybrid}. The main difference in this paper is that we will treat the rendezvous problem directly using the HCW model. More precisely, we will present the following contributions:
\begin{itemize}
    \item a hybrid control law for $u_z$ to stabilize the  out-of-plane dynamics (the $z$ direction);
    \item two independent hybrid control laws for $u_x$ and $u_y$ to stabilize the in-plane dynamics (the $x,y$ plane);
    \item each control law aims at defining both the correct position to perform efficient actions by the thrusters and the appropriate magnitude of the impulses;
    \item the control impulses $u_x$, $u_y$ and $u_z$ are not necessarily synchronized, to allow for more flexibility and to potentially reduce the number of impulses;
    \item the nonlinear control laws are equipped with a dwell-time dynamics to avoid fast consecutive use of the same input $u_x$, $u_y$, or $u_z$; this dwell-time property, among other things, avoids the
    Zeno phenomenon i.e., the occurrence of an infinite number of impulses in a finite interval of time.
\end{itemize}

In the HCW model (\ref{eq:HCW-a})--(\ref{eq:HCW-b}), one can identify two decoupled dynamics, referred to as in-plane (the $(x,y)$ directions) and out-of-plane dynamics (the $z$ direction). We carry out the hybrid stabilizers by first focusing 
on the $z$ (out-of-plane) direction and then show how the ensuing ideas can be followed for stabilizing also the more challenging in-plane direction.

\subsection{Hybrid Dynamical Systems framework}

The chaser's maneuver is represented as a sudden velocity change, given by $u$. As such, the governing dynamics are considered a hybrid system. A hybrid system integrates both continuous and discrete dynamical characteristics, permitting time-evolving processes and instantaneous transitions. Essentially, it represents a system capable of continuous evolution with potential for discrete jumps. Following the definition of hybrid dynamical systems from \cite{goebel2009hybrid},\cite{goebel2012hybrid}, such a system can be  formally defined as:
\begin{equation}
    \mathcal{H} = (\mathcal{C},\mathcal{D},F,G)
\end{equation}
where $n_x \in \mathbb{N}$ is the state dimension, $\mathcal{C} \subseteq \mathbb{R}^{n_x}$ is the flow set,  $\mathcal{D} \subseteq \mathbb{R}^{n_x}$ is the jump set, $F: \mathcal{C} \rightarrow \mathbb{R}^{n_x}$ is the flow map and  $G:\mathcal{D} \rightarrow \mathbb{R}^{n_x}$ the jump map. Mathematically, $\mathcal{H}$ is given by
\begin{subequations}
\begin{empheq}[left=\text{$\mathcal{H}$ : } \empheqlbrace]{align}
&\dot{\mathbf{x}} \in F(\mathbf{x}),\quad \mathbf{x} \in \mathcal{C} \\
&\mathbf{x}^+ \in G(\mathbf{x}),\quad \mathbf{x} \in \mathcal{D}
\end{empheq}
\label{dynhybrideq}
\end{subequations}

In the next section, we will use this framework to develop hybrid control laws for the spacecraft rendezvous problem.

\section{Stabilization in the $z$ direction}
\label{sec-outofplane}

When focusing on the $z$ direction, one can extract the following essential dynamics from the model 
\eqref{eq:HCW-c} and \eqref{eq:ImpContz}:
\begin{align}
\label{eq:flow0}
    \begin{bmatrix} 
    \dot {r}_z\\
    \dot{v}_z\end{bmatrix}&= \begin{bmatrix}
        0 & 1\\ -n^2 &0
    \end{bmatrix} \begin{bmatrix} 
    r_z\\ {{v_z}}\end{bmatrix} \quad \mbox{ when in free motion},\\
        \begin{bmatrix} 
    {r_z^+}\\
    v_z^+\end{bmatrix} &= \begin{bmatrix} 
    r_z\\
    {v_z}\end{bmatrix}  + \begin{bmatrix}
        0 \\ 1
    \end{bmatrix}u_z \quad \mbox{ when firing the input $u_z$} . 
    \label{eq:jump0}
\end{align}
With this equation in mind, we design here a hybrid feedback controller not only deciding the selection 
of the input $u_z$ but also the triggering condition for firing the input, thereby inducing a jump as in \eqref{eq:jump0}.




The proposed hybrid feedback controller comprises two internal states, one playing the role of a logic variable $q_z \in \{-1,1\}$ and one of them being a timer variable $\tau_z \in [0,2]$ whose value is constrained to evolve in a forward invariant compact set, selected as $[0,2]$ by way of suitable scaling, for simplified notation.

By gathering the overall out-of-plane closed-loop state in a vector
$\xi_z\!:=\!\begin{bmatrix} r_z&\!\!  {v_z}&\!\! q_z &\!\! \tau_z \end{bmatrix}^{\!\top}$, 
and selecting $u_z = - v_z$, so that suitable damping of the closed-loop velocity is obtained in \eqref{eq:jump0},
we may write the closed-loop hybrid dynamics as 
\begin{subequations}
\label{eq:closed_loop_z}
\begin{align}
\left\{
\begin{array}{l}
    \begin{bmatrix} 
    \dot {r}_z\\
    \dot{v}_z \end{bmatrix}
       = \begin{bmatrix}
        0 & 1\\ -n^2 &0
    \end{bmatrix} \begin{bmatrix} 
    r_z\\ {{v_z}}\end{bmatrix} ,  \\
\dot \tau_z  = \frac{n}{2\pi} (1-\text{dz}(\tau_z)), \\
\dot q_z = 0,
\end{array} \right.
\quad  & \xi_z\in \mathcal C_z,\\
\left\{\begin{array}{l}
       \begin{bmatrix} 
    {r_z^+}\\
    v_z^+\end{bmatrix} = \begin{bmatrix} 
    r_z\\
    {v_z}\end{bmatrix}  - \begin{bmatrix}
        0 \\ 1
    \end{bmatrix}\text{sat}(v_z), \\
    \tau_z^+ = 0,\\
    q_z^+ = -q,
\end{array}\right.  \quad & \xi_z\in \mathcal D_z,
\end{align}
\end{subequations}
where one clearly sees that the timer $\tau_z$ is reset to zero at each firing of the input $u_z = -\sat(v_z)$ and that it keeps track of each revolution of the oscillatory free (continuous) dynamics \eqref{eq:flow0}. In particular, after one revolution, unless it is reset before, the timer reaches the value $\tau_z = 1$. The dead-zone function, defined by  $\text{dz}(u)=u-\sat_1(u)$ with $\sat_1$ being the symmetric saturation function with maximal amplitude $1$, appearing in the flow dynamics of $\tau_z$ ensures that after $\tau_z \geq 1$, its speed is suitably slowed down until it is completely stopped once $\tau_z = 2$. This ensures that the set $[0,2]$ is forward invariant for $\tau_z$, even though a consequence of this is that once $\tau_z \geq 1$ it does not anymore keep track of the revolutions of the continuous dynamics.
About the logic variable $q_z$, note that it toggles between $-1$ and $1$ at jumps.
We may now define the
 the flow and jump sets as given by
    \begin{align}
    \label{eq:Zflowjumpsets}
         \nonumber\!\mathcal C_z  &\!= \!\! \left\{\!\xi_z\!\in\!\mathbb R^3, r_z({v_z}\!-\!nr_z)\leq 0 \mbox{ or } q_z{v_z} \leq 0 \mbox{ or } \tau_z \leq \!\tau_z^M \right\}\!, \\
        \! \mathcal D_z   &\!=\!\! \left\{\!\xi_z\!\in\!\mathbb R^3, r_z({v_z}\!-\!nr_z) \geq 0 \mbox{ and } q_z{v_z} \geq 0 \mbox{ and }\tau_z\! \geq \!\tau_z^M  \!\right\}\!,
    \end{align}
where $\tau_z^M$ is in $(0,2]$.
    The objective is to guarantee global asymptotic stability of the following set
\begin{equation}
\label{eq:Az}
    \mathcal A_z:=\left\{\xi_z: \; r_z=v_z=0, q_z\in\{-1,1\},\tau_z \in[0,2]\right\}
\end{equation}
for the hybrid closed loop, as clarified din the next statement.

\begin{theorem} \label{thm:zplane}
    The set $\mathcal A_z$ in \eqref{eq:Az} is globally asymptotically stable for the hybrid dynamics \eqref{eq:closed_loop_z}.
\end{theorem}

\begin{proof}
Consider the Lyapunov function given by
    \begin{align}
        \label{eq:Vz}
        V(\xi_z) = n^2 r_z^2+ v_z ^2,
    \end{align}
which is positive definite and radially unbounded with respect to $\mathcal A_z$. Differentiating the Lyapunov function $V$ along the solutions evolving in $\mathcal C_z$, we get that
\begin{align}
\dot V(\xi_z) = 2n^2 r_z\dot r_z+ 2 {v_z}\dot v_z= 2n^2 r_z{v_z} - 2 n^2{v_z} r_z=0.
\label{eq:Vz_flows}
\end{align}

Let us now evaluate the evolution of the Lyapunov function across jumps. 
To this end, first note that within the jump set we have $|v_z||r_z|\geq n|r_z|^2$, which implies
$|v_z| \geq n|r_z|$ and also $v_z^2 \geq n^2r_z^2$. Following the jump dynamics of the closed loop \eqref{eq:closed_loop_z}, 
it holds that
$$\begin{array}{lcl}
\Delta V(\bar z) &=&({(n r_z^+)}^2+{(v_z^+)}^2)- (n^2r_z^2+v_z^2)\\
&=&n^2{r_z}^2+({v_z}-\phi(v_z))^2-n^2{r_z}^2- v_z^2\\
&=& -\sat(v_z)(2v_z-\sat(v_z)).\\
\end{array}
$$
Since the $[0,1]$ sector property of the saturation implies $({v_z}-\sat({v_z})) \sat({v_z})\geq0 $ for all ${v_z} \neq0$, then we have
\begin{align}
\Delta V(\xi_z) \leq - {v_z} \sat({v_z}) <  0, \quad \forall \xi_z \in \mathcal D_z \setminus {\mathcal A}_z,
\label{eq:Vz_jumps}
\end{align}
where negativity comes from the fact that $v_z$ is never zero in $\mathcal D_z\setminus {\mathcal A}_z$.
Together with positive definiteness and radial unboundedness of $V$, \eqref{eq:Vz_flows} and \eqref{eq:Vz_jumps}
imply Lyapunov stability of ${\mathcal A}_z$. To prove attractivity, we use the La Salle invariance principle in \cite[Thm 1]{SeuretTAC19}. Due to the differentiability of $V$, all of the requirements on $V$ in \cite[Thm 1]{SeuretTAC19} hold true and it remains to prove that no complete solution $\psi$ exists, along which $V$ is constant and nonzero. This fact is easily proven by the property that solutions flowing with $V(\psi)>0$ also satisfy $V(\psi)>0$ and since this value remains constant (by the assumed properties of $\psi$), the solution must reach in finite time the jump set $\mathcal D_z$ where a jump is necessary to prevent flowing away from $\mathcal C_z$. Due to \eqref{eq:Vz_jumps}, $V$ must decrease across that jump, thus 
allowing us to conclude global asymptotic stability from \cite[Thm 1]{SeuretTAC19}.
\end{proof}

\section{Stabilization in the $(x,y)$ plane}\label{sec-inplane}

In this section, the objective is to follow a similar path to the one laid down in Section~\ref{sec-outofplane} for the $z$ direction, to solve the more challenging problem of stabilizing the coupled dynamics in the $(x,y)$ plane. Combining these two controllers will lead to an overall hybrid stabilizing control system.

To this end, we introduce the following dynamics, describing the hybrid evolution in the $(x,y)$ plane, as a function of the impulsive inputs $u_x$ and $u_y$,
\begin{equation}
\label{eq:xy_initial}
    \begin{array}{lcllcl}
    \begin{bmatrix} 
    \dot {r}_x\\
    \dot{v}_x\\
    \dot r_y\\
    \dot v_y
    \end{bmatrix}&\!=&\! \underbrace{\begin{bmatrix}
        0 & 1& 0& 0\\ 
        3n^2 &0&0&2n\\
        0& 0 & 0& 1\\
        0&-2n &0&0
    \end{bmatrix}}_{=A_0} \begin{bmatrix} 
    r_x\\ v_x\\r_y\\v_y\end{bmatrix}, \mbox{ in free motion}
    \\
        \begin{bmatrix} 
    r_x^+\\
    v_x^+\\
    r_y^+\\
    v_y^+\end{bmatrix} &\!=&\! \begin{bmatrix} 
    r_x\\
    v_x\\
    r_y\\
    v_y\end{bmatrix}  \!+ \!\underbrace{\begin{bmatrix}
        0 &0\\ 1&0\\ 0&0\\ 0&1
\end{bmatrix}}_{=B_0}\begin{bmatrix}
        \sat(u_x)\\
        \sat(u_y)
    \end{bmatrix}, \mbox{ when firing inputs}
    \end{array}
\end{equation}

For obtaining an insightful representation of dynamics \eqref{eq:xy_initial}, 
introduce the coordinate transformation 
\begin{equation}\label{def:tildexy}
\zeta=\begin{bmatrix}
x\\
y \\
\alpha\\
\beta
\end{bmatrix} := 
\underbrace{\begin{bmatrix}
    -3& 0& 0& -2/n\\
    0 &1& 0& 0\\ 
    0 &-2/n &1 &0\\
    -6n& 0& 0 &-3
\end{bmatrix}}_{:=T}\begin{bmatrix}
r_x\\
v_x\\
r_y\\
v_y
\end{bmatrix} 
\end{equation}

With these new coordinates, the dynamics  \eqref{eq:xy_initial} writes
\begin{equation}\label{Hxy}
    \begin{array}{lcllcl}
    \dot \zeta&=& A\zeta,\qquad & \mbox{in free motion} \\
    \zeta^+ &=& \zeta+B\begin{bmatrix}
        \sat(u_x)\\
        \sat(u_y)
    \end{bmatrix}, \qquad &  \mbox{when firing inputs}
    \end{array}
\end{equation}
with $A=TA_0T^{-1}$ and $B=TB_0$, corresponding to 
$$
A=\begin{bmatrix}
        0 & 1& 0& 0\\ 
        -n^2 &0&0&0\\
        0& 0 & 0& 1\\
        0&0 &0&0
    \end{bmatrix},\ B=\begin{bmatrix}
        0&-2/n\\ 
        1&0\\ 
        -2/n&0\\ 
        0&-3
    \end{bmatrix}
$$

The structure of matrices $A$ and $B$ highlights that, along flowing solutions, the dynamics of the system is driven by two independent subsystems, one being an oscillator (having states $x,  y$) and a second one being a double integrator  (having states $\alpha,  \beta$). Across jumps, the structure of $B$ introduces a coupling between both dynamics. Indeed, each impulsive action through $u_x$ and $u_y$ affects both the oscillator and the double integrator. Indeed, $u_x$ affects $y$ and $\alpha$, while $u_y$ affects $y$ and $\beta$. 

To simplify the controller design, we exploit the structure of the system to provide a hierarchical control law to split the action of the two control inputs. More specifically, first the control input $u_y$ is used to ensure finite-time convergence to zero of $\beta$. Then, the control input $u_x$ is used to stabilize the remaining states $[x\ y\ \alpha]^\top$, following the same type of control laws as for the dynamics in the $z$ direction, discussed in Section~\ref{sec-outofplane}. 
These two stabilizers are discussed in the next sections.

\subsection{Finite-time stabilization of $\beta$}

The proposed hybrid feedback controller for the dynamics of $\beta$ comprises one internal state, being a timer variable $\tau_\beta \in [0,2]$ whose value is constrained to evolve in a forward invariant compact set, selected as $[0,2]$ by way of a suitable scaling, for simplified notation.

By gathering the plant-controller state in a vector
$\xi_\beta:=\begin{bmatrix} \beta&  \tau_\beta \end{bmatrix}^\top$, 
and selecting $u_y = -\beta/3$, we may write the closed loop hybrid dynamics as follows
\begin{subequations}
\label{eq:closed_loop_beta}
\begin{align}
\left\{
\begin{array}{l}
    \dot {\beta}=0; \\
\dot \tau_\beta  = \frac{n}{2\pi} 
(1-\text{dz}(\tau_\beta)) \\
\end{array} \right.
\quad  & \xi_\beta\in \mathcal C_\beta\\
\left\{\begin{array}{l}
    {\beta^+}=\beta -3 \sat(\beta/3) \\
    \tau_\beta^+ = 0,\\
\end{array}\right.  \quad & \xi_\beta\in \mathcal D_\beta.
\end{align}
\end{subequations}

Mimicking the solution in Section~\ref{sec-outofplane},
 this hybrid dynamics includes a timer $\tau_\beta$, which is reset to zero at each firing of the input $u_y = \sat(3\beta)$ and obeys flow dynamics ensuring that $\tau_\beta$ never leaves the compact set $[0,2]$.

For a given $\tau_\beta^M$ in $(0,2]$, we may then define the flow and jump sets as follows
    \begin{align}
    \label{eq:Betaflowjumpsets}
         \mathcal C_\beta  =  \left\{\xi_\beta\in\mathbb R^2,\; \tau_\beta \leq \tau_\beta^M \right\},\ 
         \mathcal D_\beta   =  \left\{\xi_\beta\in\mathbb R^2,\; \tau_\beta \geq \tau_\beta^M  \right\}\!,
    \end{align}

The objective is now to guarantee global finite-time stability of the following set
\begin{equation}
\label{eq:Abeta}
    \mathcal A_\beta:=\left\{\xi_\beta: \; \beta=0,\;\tau_\beta \in[0,2]\right\}
\end{equation}
for the hybrid closed loop, as clarified in the next statement.
We recall that finite-time stability comprises global asymptotic stability and finite-time convergence.

\begin{theorem} \label{thm:beta}
 The set $\mathcal A_\beta$ in \eqref{eq:Abeta} is finite-time stable  for the hybrid dynamics \eqref{eq:closed_loop_beta}.
\end{theorem}

\begin{proof}
Consider the Lyapunov function given by $V(\xi_\beta)=\beta^2$.
The derivative of $V$ along flowing trajectories is zero (because $\dot \beta = 0$) and, its variation across jumps writes
$$\begin{array}{lcl}
\Delta V(\xi_\beta)&=&\left(\beta-3\sat\left(\frac{\beta}{3}\right)\right)^2-\beta^2\\
&=& 6\sat\left(\frac{\beta}{3}\right)\left(\sat\left(\frac{\beta}{3}\right)-\frac{\beta}{3}\right)-\sat\left(\frac{\beta}{3}\right)\frac{\beta}{3}
\end{array}$$
Since the $[0,1]$ sector property of the saturation implies $({\beta/3}-\sat({\beta/3})) \sat({\beta/3})\geq0 $ for all ${\beta} \neq0$, then we have
$$\begin{array}{lcl}
\Delta V(\xi_\beta)&\leq&-\sat\left(\frac{\beta}{3}\right)\frac{\beta}{3}<0,\ \forall \xi_\beta\in\mathcal D_\beta\setminus \mathcal A_\beta.
\end{array}$$
Such a decrease condition, together with the persistent jumping properties of the closed loop \eqref{eq:closed_loop_beta}, implies global asymptotic stability of $\mathcal A_\beta$.

To prove finite-time convergence, please note that, at each jump, two cases may occur. If $\beta$ is not in $[-3\umax,3\umax]$, then $|\beta^+|=|\beta|-3\umax\neq 0$. Eventually, after a finite number of jumps, $\beta$ will enter the interval $[-3\umax,3\umax]$. In this situation $\sat(\beta/3)=\beta/3$ and $\beta^+=0$, which proves finite-time convergence and concludes the proof. 
\end{proof}

\begin{remark}
Note that the definition of the flow and jump sets $\mathcal C_\beta $ and $\mathcal D_\beta $ imposes periodic impulses. Such a periodic behavior could have been modelled using a simple timer instead of this more involved solution. This choice is made to be consistent with the other timers required for the two other hybrid controller. 
\end{remark}

\begin{remark}
Even though the control input $u_y$ is triggered periodically, the finite convergence of $\beta$ to the origin ensures that the magnitude of the control law $\sat(\beta/3)$ will also be zero in finite time, and no control action will be applied. 
\end{remark}

\subsection{Finite-time stabilization of $x,y,\alpha$}

Following the previous paragraph, a first layer of the control law suitably selects the
control input $u_y$ to ensure
finite-time convergence to zero of the variable $\beta$, in addition to finite-time convergence to zero of $u_y$ itself. In this section, we will exploit a suitable hybrid feedback selection of $u_x$ to stabilize the variables $x,y$ and $\alpha$ to the origin. Once $\beta$ and $u_y$ have converged to zero, 
we may design a feedback controller based on a logic variable $q_\alpha\in \{-1,1\}$ and on a time variable $\tau_\alpha \in [0,2]$, by focusing on the following
 reduced dynamics with plant states $[x\ y\ \alpha]^\top$ and controllers states $q_\alpha$ and $\tau_\alpha$ gathered in an overall vector $\xi_\alpha$,
\begin{subequations}
\label{eq:closed_loop_alpha}
\begin{align}
\left\{
\begin{array}{l}
    \begin{bmatrix}
      \dot   x\\
      \dot   y\\
       \dot  \alpha
    \end{bmatrix}
     = \begin{bmatrix}
        0 & 1& 0\\ 
        -n^2 &0&0\\
        0& 0 & 0\\
    \end{bmatrix}\begin{bmatrix}
         x\\
         y\\
         \alpha
    \end{bmatrix} \\
\dot q_\alpha=0\\
\dot \tau_\alpha =\frac{n}{2\pi} (1-\text{dz}(\tau_\alpha)) 
\end{array} \right.
\quad & \xi_\alpha
\in \mathcal C_\alpha\\
\nonumber\\
\left\{\begin{array}{l}
    \begin{bmatrix}
         x^+\\
         y^+\\
         \alpha^+
\end{bmatrix}=\begin{bmatrix}
         x\\
         y\\
         \alpha
    \end{bmatrix} +\begin{bmatrix}
        0\\ 
        1\\ 
        -\frac{2}{n}\\ 
    \end{bmatrix}\sat(u_x) \\
    q_\alpha^+=-q_\alpha\\
    \tau_\alpha^+ = 0,\\
\end{array}\right.  \quad & \xi_\alpha
\in \mathcal D_\alpha,
\end{align}
\end{subequations}

Following the control design of the dynamics in the $z$ direction (see Section~\ref{sec-outofplane},
we use the following hybrid control law for \eqref{eq:closed_loop_alpha}, consisting in the control input $u_x= \frac{n\alpha}{4}-\frac{y}{2}$, with 
\begin{equation}
 \begin{array}{lcl}    
    \mathcal C_\alpha &=&
    \left\{ \xi_\alpha \in\mathbb R^3, 
    \begin{array}{lr}
  &(y-\frac{n}{2}\alpha-nx)x \leq 0,\\  \mbox{ or } &q_\alpha (y-\frac{n}{2}\alpha)\leq 0,\\
   \mbox{ or }& \tau_\alpha
   \leq\tau_\alpha^M
    \end{array}
    \right\},\\
\\
   \mathcal D_\alpha &=&\left\{ \xi_\alpha \in\mathbb R^3, \begin{array}{lr}
 &(y-\frac{n}{2}\alpha-nx)x \geq 0,\\
 \mbox{ and } &q_\alpha (y-\frac{n}{2}\alpha)\geq 0,\\
   \mbox{ and }& \tau_\alpha
   \geq\tau_\alpha^M.
    \end{array}
   \right\},\\
   \end{array}
\end{equation}
for a given $\tau_\alpha^M$ in $(0,2]$.

    
This feedback system ensures global asymptotic stability of the following set
\begin{equation}
\label{eq:Aalpha}
    \mathcal A_\alpha:=\left\{\xi_\alpha \in\mathbb R^3:\; \xi_\alpha =0,\; q_\alpha \in\{-1,1\},\;\tau_\alpha \in [0,2]\right\},
\end{equation} 
as clarified din the next statement.

\begin{theorem}\label{thm:xyplane}
    The set $\mathcal A_\alpha$ in \eqref{eq:Aalpha} is globally asymptotically stable for the hybrid dynamics \eqref{eq:closed_loop_alpha}.
\end{theorem}

\begin{proof}
Let us consider the quadratic Lyapunov function
$$V_\alpha(\xi_\alpha)=n^2  x^2 +   y^2 +\frac{n^2}{4}  \alpha^2.$$


Computing the directional derivative of $V_\alpha$ along the flow map in \eqref{eq:closed_loop_alpha} yields: 
$$\begin{array}{lcl}
\dot V_\alpha(\xi_\alpha)&=&2n^2  x \dot{  x} +2  y\dot{  y} +2\frac{n^2}{4}  \alpha\dot{  \alpha}\\
&=&2n^2 x  y -2n^2 y x  +2\frac{n^2}{4} \alpha \cdot 0=0\\
\end{array}
$$
Hence, the Lyapunov function $V_\alpha$ is constant along flowing solutions.

Let us now evaluate the variation of the Lyapunov across jumps. following the same steps as in the proof for the $z$ direction, let  
us first note that within the jump set we have $|y-n\alpha/2||x|\geq n|x|^2$, which implies $|y-n\alpha/2| \geq n|x|$ and, in particular, 
\begin{equation}\label{ineq:yalpha_x}
    (y-n\alpha/2)^2 \geq n^2x^2.
\end{equation}

Computing now the variation of the Lyapunov function across jumps yields
$$\begin{array}{lcl}
\Delta V_\alpha(\xi_\alpha)&\!\!\!\!\!=&\!\!\!\!(n  x^+)^2 + (  y^+)^2  +\frac{n^2}{4}(  \alpha^+) ^2\\
&&\!\!\!\!-n^2  x^2 -   y^2 -  \frac{n^2}{4} \alpha ^2\\
 \end{array}
$$
where we have used that the fact that $x$ is not modified across jumps. Substituting the input selections $  y^+ = y+\sat(u_x)$ and $\alpha^+= \alpha-\frac{2}\sat(u_x)$ in the expression above, we obtain
$$\begin{array}{lcl}
\Delta V_\alpha (\xi_\alpha )&=&
(y -\sat(u_x))^2 +\frac{n^2}{4}(\alpha-\frac{2}{n}\sat(u_x)) ^2\\
&&-   y^2 -   \frac{n^2}{4}\alpha ^2\\
&=&
2\sat(u_x)\left( -(y-\frac{n}{2} \alpha)+\sat(u_x) \right)\\
&=&
-2\sat(u_x)\left( 2u_x-\sat(u_x) \right)
 \end{array}
$$
Following the same argument as the one in Theorem~\ref{thm:zplane} for the $z$ direction, the term $\sat(u_x)\left(u_x-\sat(u_x) \right)$ is positive for any nonzero value of $u_x=\frac{1}{2}y-\frac{n}{4}\alpha\neq0$. This implies 
$$
\Delta V_\alpha (\xi_\alpha )\leq-2\sat(u_x)u_x<0,\quad \forall \xi_\alpha \mbox{ s.t. } u_x\neq0.\\
$$
The proof is concluded by applying the La Salle invariance principle \cite[Chapter 8]{goebel2012hybrid} by noting that the unique invariant trajectory of hybrid system \eqref{eq:closed_loop_alpha} which verifies $u_x=n\alpha/4-y/2=0$ and also  $x=0$, thanks to \eqref{ineq:yalpha_x},  is the trivial solution $\xi_\alpha=0$. Indeed, the second equation implies that $y=n\alpha/2$. According to the flow equation, $\dot x=y$, the state $x$ remains at $0$ if and only $y=n\alpha/2=0$. 
\end{proof}

\section{Combined hybrid feedback}
\label{sec:combined_feedback}

In Sections~\ref{sec-outofplane} and~\ref{sec-inplane} we have designed three nested hybrid controllers, each of them inducing desirable properties of the corresponding dynamics. Their hybrid combination is discussed here, where we illustrate the use of reduction arguments to prove global asymptotic stability of the origin for the overall control law. 

To combine the three controllers, let us 
first define the overall state 
$\xi = [\xi_z^\top \ \xi_\beta^\top\ \xi_\alpha^\top]^\top$
for the closed-loop dynamics:
\begin{align}
    \xi = [r_z\ v_z\ \tau_z\ q_z\ \beta\ \tau_\beta\ x\ y\ \alpha\ q_\alpha\  \tau_\alpha]^\top.
\end{align}
State $\xi$ clearly evolves in the following set:
\begin{align}
{\mathbb X} = {\mathbb R}^2 \times [0,2] \times {\mathcal Q} \times
{\mathbb R} \times [0,2] \times {\mathbb R}^3 \times {\mathcal Q} \times [0,2],
\end{align}
where we denoted ${\mathcal Q} = \{-1,1\}$.

Based on state $\xi$
we may define the following extended selections of the impulsive feedback control law:
\begin{align}
\kappa_z(\xi) = \begin{bmatrix}
    0 \\ 0 \\ u_z
\end{bmatrix};\quad
\kappa_\beta(\xi) = \begin{bmatrix}
    0 \\ u_y \\ 0
\end{bmatrix};\quad
\kappa_\alpha(\xi) = \begin{bmatrix}
    u_x \\ 0 \\ 0
\end{bmatrix},
\end{align}
with $u_z = -v_z$, $u_y = -\beta/3$ and $u_x = \frac{n\alpha}{4} - \frac{y}{2}$ as per the selections in \eqref{eq:closed_loop_z}, \eqref{eq:closed_loop_beta}, and \eqref{eq:closed_loop_alpha}.

We may then just as well define the three following jump maps and jump sets, 
each of them characterizing the corresponding stabilizer, whose properties have been established in Theorems~\ref{thm:zplane},~\ref{thm:beta} and~\ref{thm:xyplane}:
\begin{align*}
g_z(\xi) &= \begin{bmatrix}
    r \\ v + \sat(\kappa_z(\xi))
\end{bmatrix}, \quad \xi\in \overline{\mathcal{D}}_z:= \{\xi: \xi_z \in {\mathcal D}_z\} \\
g_\beta(\xi) &= \begin{bmatrix}
    r \\ v + \sat(\kappa_\beta(\xi))
\end{bmatrix}, \quad \xi\in \overline{\mathcal{D}}_\beta:= \{\xi: \xi_\beta \in {\mathcal D}_\beta\} \\
g_\alpha(\xi) &= \begin{bmatrix}
    r \\ v + \sat(\kappa_\alpha(\xi))
\end{bmatrix}, \quad \xi\in \overline{\mathcal{D}}_\alpha:= \{\xi: \xi_\alpha \in {\mathcal D}_\alpha\}.
\end{align*}

The overall control scheme is then the one prioritizing jumps, accounting for the three selections above. In particular, we may select the jump set ${\mathcal D}\subset {\mathbb X}$ and the flow set  ${\mathcal C}\subset {\mathbb X}$ as
\begin{align}
    {\mathcal D} = \overline{\mathcal{D}}_z \cup \overline{\mathcal{D}}_\beta \cup \overline{\mathcal{D}}_\alpha, \quad
    {\mathcal C} = \overline{{\mathbb X} \setminus    {\mathcal D}},
\end{align}
where the overline denotes the closure operation, so that the flow set is the closed complement of the jump set.

As for the jump map, following standard practices, we select it as a set-valued jump map $G$ whose graph is the union of the graphs of $g_z$, $g_\beta$ and $g_\alpha$ defined above.
More specifically, $G$ contains all the possible update laws included in the three pairs $(\overline{\mathcal{D}}_z,g_z)$, $(\overline{\mathcal{D}}_\beta,g_\beta)$, and $(\overline{\mathcal{D}}_\alpha,g_\alpha)$ so that, as an example, 
when $\xi \in (\overline{\mathcal{D}}_z \cap \overline{\mathcal{D}}_\beta) \setminus \overline{\mathcal{D}}_\alpha$, then $G(\xi) = \{g_z(\xi)\} \cup \{g_\beta(\xi)\}$, and similarly for the other cases.

Denoting by $f(x)$ the juxtaposition of the flow maps in \eqref{eq:closed_loop_z}, \eqref{eq:closed_loop_beta} and \eqref{eq:closed_loop_alpha}, respectively, we may write the overall closed loop as:
\begin{subequations}
\label{eq:closed_loop_overall}
\begin{align}
\dot \xi &= f(\xi), && \xi \in {\mathcal{C}}\\
\xi^+ &\in G(\xi), && \xi \in {\mathcal{D}}.
\end{align}
\end{subequations}
The following result then states and proofs our main result about stability properties of the following set for the closed-loop dynamics \eqref{eq:closed_loop_overall},
\begin{align}
\label{eq:attractor}
{\mathcal A}= \{\xi: r_z=v_z=0, x=y=0, \alpha=\beta=0 \}.
\end{align}

\begin{theorem}
The set ${\mathcal A}$ is globally asymptotically stable for the overall closed-loop system \eqref{eq:closed_loop_overall}.
\end{theorem}

\begin{proof}
To begin with, by construction, the hybrid dynamics \eqref{eq:closed_loop_overall} satisfies the hybrid basic conditions in \cite[As. 6.5]{goebel2012hybrid} because sets ${\mathcal C}$ and ${\mathcal D}$ are closed, function $f$ is continuous, and map $G$ is outer semi-continuous and locally bounded by construction. 

Then observe that, due to the presence of the three timers $\tau_z$, $\tau_\alpha$ and $\tau_\beta$, which are reset to zero across jumps to $g_z(\xi)$, $g_\beta(\xi)$, and $g_\alpha(\xi)$, respectively, all closed-loop solutions present a bounded number of jumps in each compact time interval $[0,T]$ so that all solutions exhibit unbounded domains in the $t$ direction. 

Then note that the dynamics of the states $\xi_z$ and $\xi_{\beta}$ are completely decoupled and, due to Theorems~\ref{thm:zplane} and~\ref{thm:beta}, they are such that the set 
\begin{align}
    \Gamma := \{\xi \in {\mathbb X}: r_z=v_z=0, \beta=0\}
\end{align}
is globally asymptotically stable. This result is a straightforward consequence of using the function $V(\xi_z,\xi_\beta) = n^2 r_z^2 + v_z^2 + \beta^2$, combining the two functions used in Theorems~\ref{thm:zplane} and~\ref{thm:beta}. Moreover, due to Theorem~\ref{thm:beta}, $\beta$ converges to zero in finite time.

\begin{figure*}
\begin{center}
\includegraphics[width=1\columnwidth]{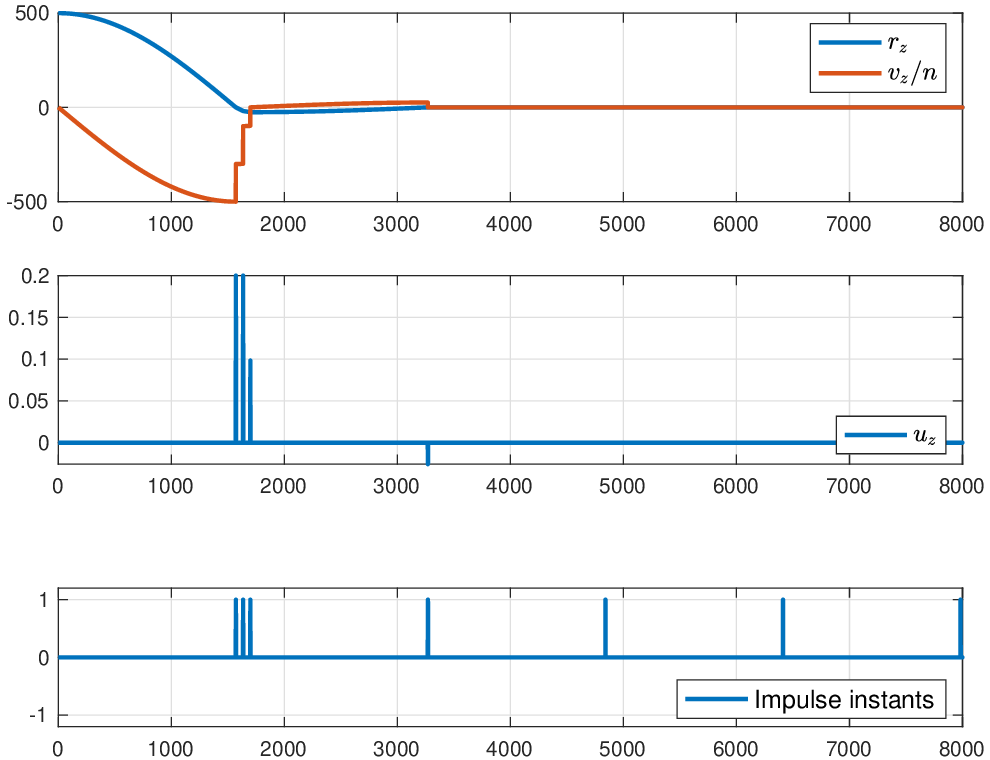}  \includegraphics[width=1.\columnwidth]{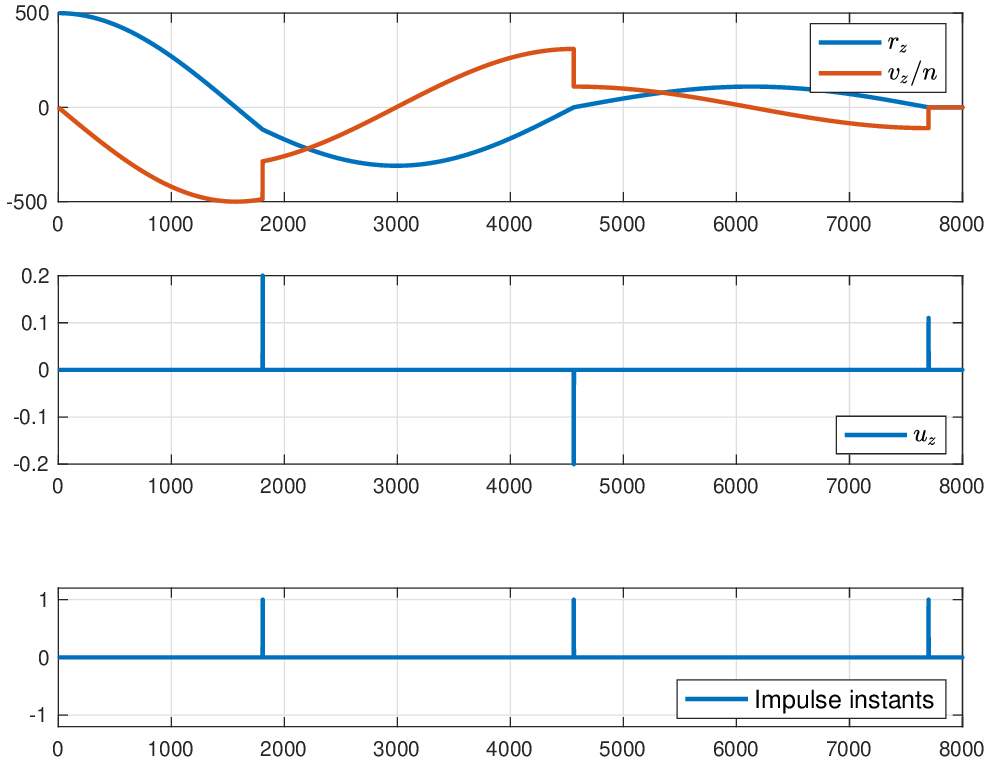}   
\caption{Simulation of system \eqref{eq:closed_loop_z} with $\tau_z^M=0.01$ (left) and $\tau_z^M=0.25$ (right). The figure shows the evolution of the state variables $r_z,v_z$ (top), the magnitude of the control input (middle) and the instants of impulses (bottom).}
\label{fig:z}
\end{center}
\end{figure*}

Consider now the result of Theorem~\ref{thm:xyplane} and note that it correspond to showing that when restricting the dynamics to initial conditions in the set $\Gamma$ (where $\beta=0$ and $u_y=0$), the set ${\mathcal A}$ is globally asymptotically stable. Using the notation in \cite{MaggioreTAC18},  we may say that ${\mathcal A}$ is globally asymptotically stable relative to $\Gamma$. 
Within the above setting, we may apply \cite[Cor. 4.8]{MaggioreTAC18}, because the two stability results above imply that items i') and ii') of that corollary are satisfied with $\Gamma_1 ={\mathcal A}$ and $\Gamma_2 = \Gamma$. Then the corollary implies asymptotic stability of ${\mathcal A}$ with basin of attraction coinciding with the set of initial conditions from which all solutions are bounded. Since $\beta$ converges to zero in finite time, boundedness of all solutions follows from the fact that: (i) the components of $\xi_z$ and $\xi_\beta$ are bounded due to global asymptotic stability of $\Gamma$; (ii) the components of $\xi_\alpha$ are bounded because they experience an initial transient 
(which is clearly bounded) within the time interval
before time $t+j = T$ when $\beta(t,j)$ converges to zero. Then, after that time $T$, the asymptotic stability bound established in Theorem~\ref{thm:xyplane} implies boundedness of the tail of $\xi_\alpha$. As a conclusion, all solutions are bounded, and global asymptotic stability of ${\mathcal A}$ follows from 
\cite[Cor. 4.8]{MaggioreTAC18}.
\end{proof}

\section{Simulation results}\label{sec-simulations}

This section aims to illustrate the numerical application of the proposed control laws. In all simulations, a symmetric saturation has been applied to limit the magnitude of the impulses. The saturation level is chosen as $\umax=0.2m/s$.

Figure \ref{fig:z} presents two simulations for the closed-loop system \eqref{eq:closed_loop_z} with two different values of the dwell time parameter $\tau_z^M$. In both cases, the state variables ($r_z, v_z$) converge asymptotically to the origin. These figures also demonstrate that our control triggers impulses when $r_z$ crosses zero, which is the situation when the control action is most efficient. When the dwell-time parameter is smaller than 0.25, equivalent to a quarter of a rotation, an additional impulse is triggered to compensate for the saturation. Both cases show a trade-off between the rate of convergence (faster with $\tau_z^M=0.01$) and the cost of consumption (lower with $\tau_z^M=0.25$, i.e., only three impulses). 

\begin{figure}
    \begin{center}
\includegraphics[width=1.\columnwidth]{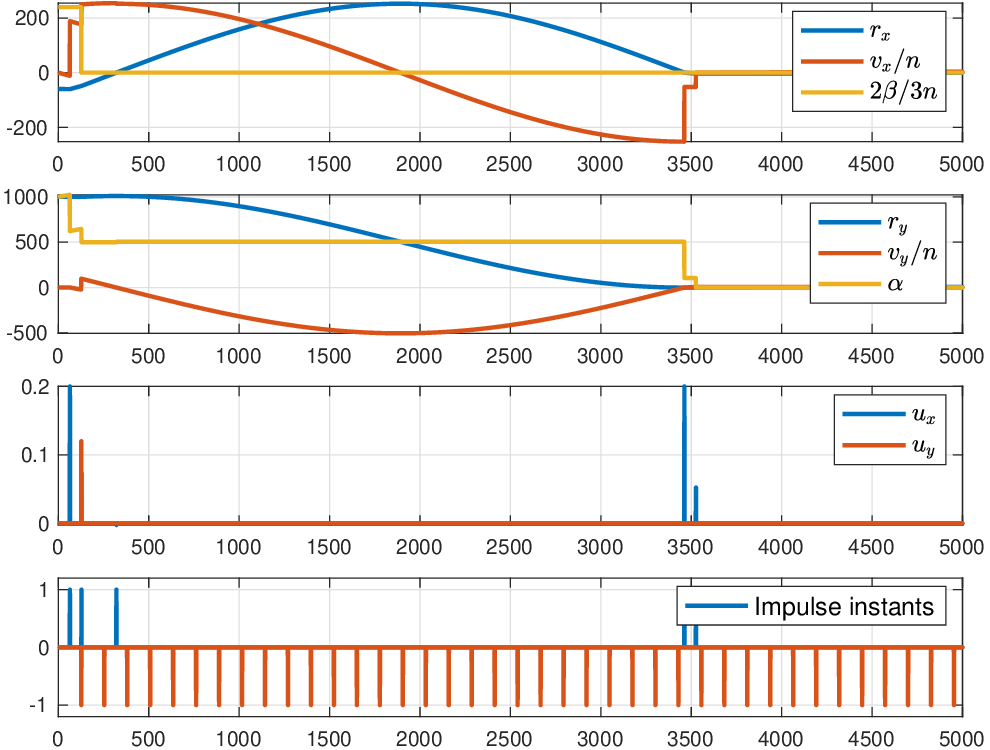}
\caption{Evolution in the original coordinates $r_x,v_x/n$ together with the transformed state $2\beta/3$ from \eqref{def:tildexy} (top plot), as well as the original coordinates $r_y,v_y/n$ together with the transformed state $\alpha$ (second plot). Magnitude of the control input (third plot) and instants of impulses (bottom plot).}
\label{fig:plot_gen_xy}
\end{center}
\end{figure}

\begin{figure}
\begin{center}
\includegraphics[width=1.\columnwidth]{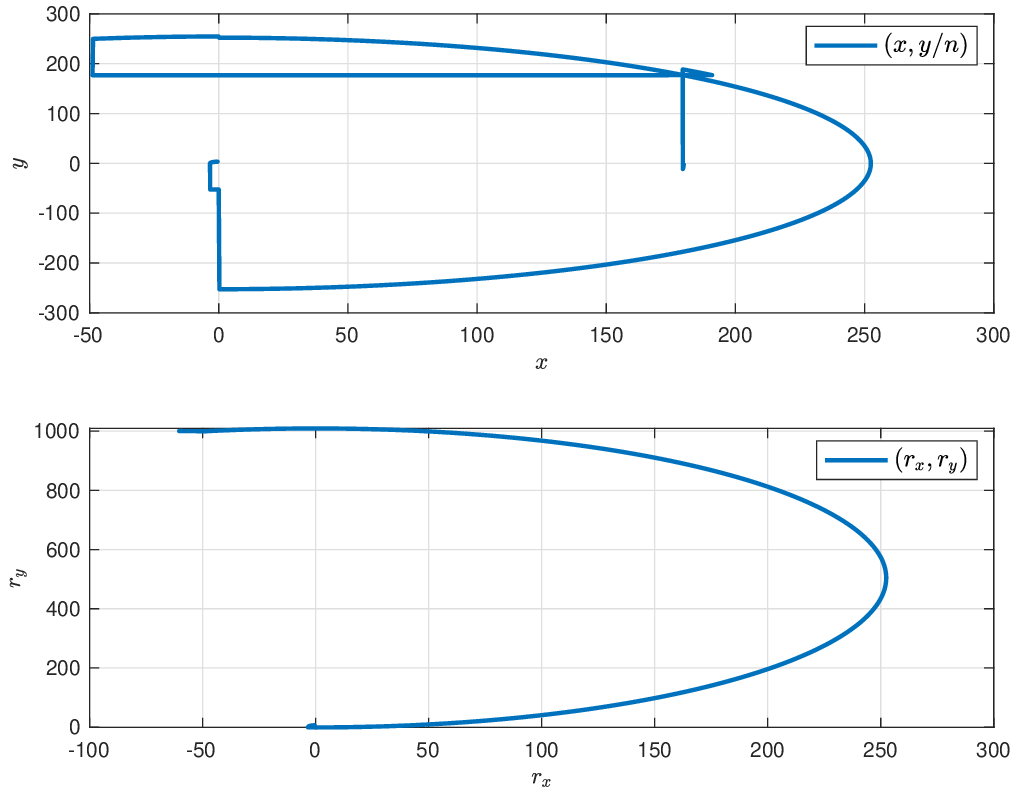}   
\caption{Evolution of the state variables $x,y/n$ (top) i.e. in the transformed coordinates $\zeta$ of \eqref{def:tildexy}, and the resulting trajectories of the original coordinates ($r_x,r_y)$ (bottom).}
\label{fig:plot_phase_xy}
\end{center}
\end{figure}

Figures \ref{fig:plot_gen_xy} and \ref{fig:plot_phase_xy} depict a simulation of the system \eqref{eq:closed_loop_beta}, \eqref{eq:closed_loop_alpha} (in-plane) with the dwell-time parameters $\tau_\alpha^M=0.01$ and $\tau_\beta^M=0.02$, along with the initial condition $[-60\ 0\ 1000\ 0]^\top$. This simulation provides an overview of the entire dynamics in the $(x, y)$-plane, which includes the closed-loop systems \eqref{eq:closed_loop_beta} (associated with state $\xi_\beta$) and \eqref{eq:closed_loop_alpha} (associated with state $\xi_\alpha$).
The bottom plot in Figure \ref{fig:plot_gen_xy} shows the periodic impulses (in red) generated by system \eqref{eq:closed_loop_beta}. As mentioned earlier in the paper, although the system regularly enters the jump set, no control action is required after a while, since the variable $\beta$ reaches zero after only one impulse (see the red impulse $u_y$ in the third plot).

The two graphs at the top of Figure \ref{fig:plot_gen_xy} demonstrate that the trajectory of the closed-loop systems \eqref{eq:closed_loop_beta}, \eqref{eq:closed_loop_alpha} converges asymptotically to the origin. Since the dwell time parameter $\tau_\alpha^M$ is less than 0.25, the control law allows for a series of three successive impulses to compensate for the effects of saturation at the maximum amplitude $\umax$ and achieve a fast convergence rate. As for the $z$ axis, selecting larger values of $\tau_\alpha^M$ (larger than 0.25), it is possible to reduce the number of impulses (consumption), but at the cost of deteriorating the convergence rate.

Finally, Figure \ref{fig:plot_phase_xy} illustrates the trajectory of the closed-loop systems in the $(r_x, r_y)$ plane (bottom). The seemingly simple behavior of the resulting solution hides a non-trivial hybrid trajectory of variables ($x, y$) depicted in the top graph, which experiences several jumps throughout the simulation.

\section{Conclusion}\label{sec-concl}
This paper presented a hybrid dynamical system approach for the impulsive control in spacecraft rendezvous and proximity operations, using the Hill-Clohessy-Wiltshire model. The control design problem was tackled by separating the out-of-plane from the in-plane dynamics, with a distinct feedback control law developed for each one of them. These laws were grounded on specially designed Lyapunov functions, accounting for thrusters saturation. Future work will include addressing minimum impulse bit requirements, in addition to safety constraints and the use of these control laws in more complex rendezvous scenarios, such as formation-flying (multiple spacecraft rendezvous), eccentric orbits (time-varying dynamics) and Halo orbit rendezvous (highly nonlinear dynamics). In addition the performance our hybrid control design should be compared and evaluated with other approaches (e.g. MPC) in a high-fidelity simulator.

\section*{ACKNOWLEDGMENT}
The work of A. Seuret was supported by ``European Union NextGenerationEU" and by the Spanish Agency for Research (AEI) through the ATRAE grant ATR2023-145067.
R. Vazquez was supported by grant TED2021-132099B-C33 funded by MCIN/ AEI/ 10.13039 /501100011033 and by ``European Union NextGenerationEU/PRTR''.


\bibliography{Rendezvous}
\bibliographystyle{ieeetr}


\end{document}